\documentclass[runningheads]{llncs}
\usepackage[T1]{fontenc}
\usepackage{graphicx,verbatim}
\begin{document}
\title{UnPuzzle: A Unified Framework for Pathology Image Analysis}
%

\author{Dankai Liao\inst{1} \and Sicheng Chen\inst{1} \and Nuwa Xi\inst{1} \and Qiaochu Xue\inst{1} \and Jieyu Li\inst{1} \and Lingxuan Hou\inst{1} \and Zeyu Liu\inst{1} \and Chang Han Low\inst{1} \and Yufeng Wu\inst{1} \and Yiling Liu\inst{1} \and Yanqin Jiang\inst{1} \and Dandan Li\inst{1} \and Shangqing Lyu\inst{1}}
\authorrunning{Dankai Liao et al.}
\institute{
    $^{1}$PuzzleAI Pte Ltd, Singapore 229594, Republic of Singapore \\
    \email{\{dankai, chensc, nuwaxi, qiaochuxue, jieyuli, lingxuanhou, zeyuliu, changhan, yufengwu, yilingwu, yanqinjiang, dandanli, shangqinglyu\}@puzzleai.sg}
}

\maketitle              
\begin{abstract}
Pathology image analysis plays a pivotal role in medical diagnosis, with deep learning techniques significantly advancing diagnostic accuracy and research. While numerous studies have been conducted to address specific pathological tasks, the lack of standardization in pre-processing methods and model/database architectures complicates fair comparisons across different approaches. This highlights the need for a unified pipeline and comprehensive benchmarks to enable consistent evaluation and accelerate research progress. In this paper, we present UnPuzzle, a novel and unified framework for pathological AI research that covers a broad range of pathology tasks with benchmark results. From high-level to low-level, upstream to downstream tasks, UnPuzzle offers a modular pipeline that encompasses data pre-processing, model composition, task configuration, and experiment conduction. Specifically, it facilitates efficient benchmarking for both Whole Slide Images (WSIs) and Region of Interest (ROI) tasks. Moreover, the framework supports various learning paradigms, including self-supervised learning, multi-task learning, and multi-modal learning, enabling comprehensive development of pathology AI models. Through extensive benchmarking across multiple datasets, we demonstrate the effectiveness of UnPuzzle in streamlining pathology AI research and promoting reproducibility. We envision UnPuzzle as a cornerstone for future advancements in pathology AI, providing a more accessible, transparent, and standardized approach to model evaluation. The UnPuzzle repository is publicly available at https://github.com/Puzzle-AI/UnPuzzle.


\end{abstract}

\section{Introduction}

Pathological image analysis plays a pivotal role in modern healthcare, where advances in deep learning have driven significant progress \cite{song2023artificial}. While most studies focus on applying AI within a single and specific topic, recent foundational models have been proposed to evaluate performance across various downstream tasks. Pretrained on large-scale datasets and applied to multiple applications, these models have significantly accelerated the integration of AI algorithms into pathology \cite{conch,uni,gigapath}.

Despite rapid progress, pathological data processing remains complex. First, researchers face challenges due to the structure of two distinct image types: Whole Slide Images (WSIs), gigapixel-scale scans covering a wide region, and Regions of Interest (ROIs), which are specific diagnostic regions akin to natural images \cite{puzzletuning}. Accordingly, the distinct characteristics of features distributed within these images introduce multiple complex stages of data processing \cite{clam,gigapath}, which necessitate multifaceted data handling, complicating AI workflows. Most studies design workflows tailored to their specific data, often lacking compatibility with each other due to variation of data format and preprocessing pipelines \cite{abmil,dsmil,clam,uni,virchow2,gigapath,musk}. Specifically, from stain standardization, magnification selection to feature embedding, the inherent complexity of pathology data workflows makes it challenging to compare models without fully replicating their entire processes. This leads to non-generalized practices and high engineering barriers, hindering reproducibility and scalability.

Moreover, achieving robust results in pathological studies generally requires both upstream pre-training and downstream fine-tuning, further complicating AI workflow design. In the upstream stage, the absence of a modular, end-to-end pipeline impedes efficient model construction. In the downstream stage, pathological diagnosis encompasses various specific high-level vision tasks, such as classification (e.g., cancer detection), regression (e.g., tumor purity estimation \cite{purity}), and low-level vision tasks (e.g., cell nuclei segmentation). Additionally, to develop a more comprehensive model, recent studies have applied more complex pipelines such as multi-task learning. Highlighting the challenge, the field calls for an open, generalized code pipeline capable of handling diverse tasks across various data types.

\begin{figure}[htbp]
  \centering
   \includegraphics[width=1.0\linewidth]{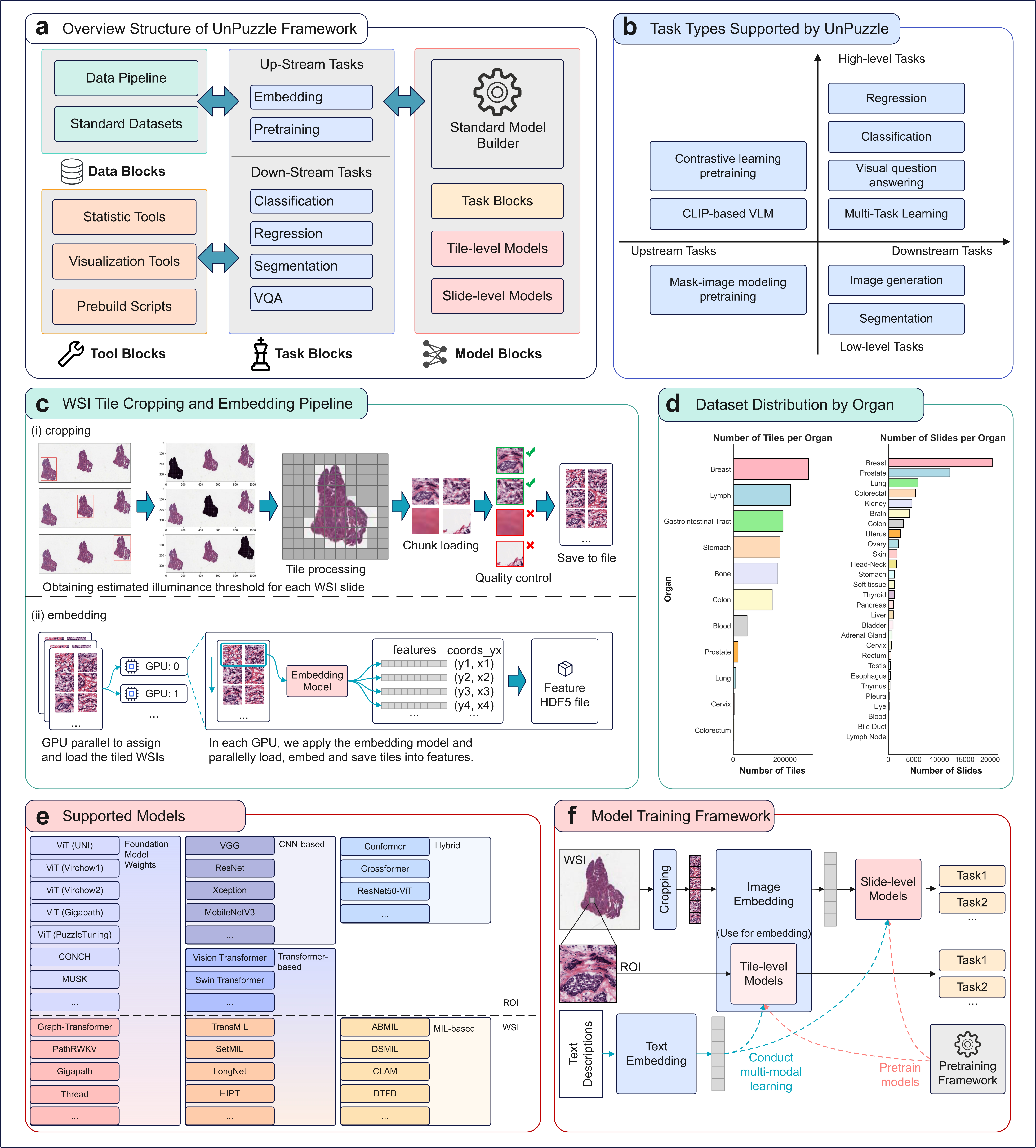}
   \caption{Structure, functions, and evaluation results of UnPuzzle framework. (a) demonstrates the overall structure of UnPuzzle framework. (b) demonstrates various tasks covered by UnPuzzle. (c) illustrates the cropping and embedding pipeline to build up the standard dataset. (d) summarizes the distribution of all collected datasets for benchmarking. (e) demonstrates the various models that UnPuzzle has covered. (f) demonstrates how UnPuzzle compose the models and task modules for pre-training and downstream tasks.}
   \label{fig:fig_large}
\end{figure}

Contributing to the community, we present UnPuzzle, one of the first end-to-end, modular, open-source frameworks designed for pathological image analysis. As a comprehensive platform, UnPuzzle standardizes data organization, preprocessing, model construction, and task evaluation workflows, streamlining the development of pathology AI models. We also release UnPuzzle with a benchmark, covering 30+ tile-level models and 20+ slide-level models, supporting pre-training frameworks and a range of downstream tasks across more than 100 datasets. By providing fair benchmarking and open-sourcing all code and results, this framework can promote transparency, accessibility, and inclusivity in pathological image analysis.

\section{Method}

The UnPuzzle framework mainly consists of 3 key components: Data blocks, Task blocks, and Model blocks, as shown in Figure~\ref{fig:fig_large}.a. First, a processed standard dataset is loaded, and then task modules for certain upstream/downstream are invoked with the foundational models to complete the model training or evaluation process.

Beyond these core components, we have developed a Browser-Server (B-S) application that facilitates model training and evaluation via a web-based user interface. The demo of this platform will be available on our website. The platform offers an intuitive and user-friendly interface for in-depth analysis of both ROIs and WSIs data, leveraging foundational models and visual-language models for enhanced functionality.

\subsection{Data Blocks}

\subsubsection{Pre-processing} 

As illustrated in Figure~\ref{fig:fig_large}.c, we design an efficient parallel processing pipeline for WSI datasets that comprises two main steps: tile cropping and tile embedding.

During the cropping stage, we load each WSI at the intended microns-per-pixel (mpp) resolution. Next, we apply Otsu’s thresholding to the entire WSI to obtain a global tissue threshold and identify unconnected tissue regions. Then, for each tissue region, chunk extraction and tile quality control are performed sequentially. To ensure tile quality, we perform two filtering steps: (1) removing tiles with insufficient tissue coverage and (2) removing tiles with low pixel variance. Lastly, all filtered tiles and their corresponding pixel locations are saved to disk.
During the embedding stage, the previously cropped tiles are processed by a GPU-accelerated embedding model. After the embedding step, the output features are stored in an HDF5 format file (with .pth embeddings labeled as “features”) and the corresponding tile location indices are saved as “coords\_xy.” This two-stage pipeline tile-cropping followed by tile-embedding enables fast and effective processing for large-scale WSI datasets while maintaining high data quality through rigorous filtering and parallelization.

\subsubsection{Datasets and task configuration design} 

To facilitate a range of downstream tasks for each dataset, we designed a standard dataset format. The dataset is organized into two main parts: (1) a set of WSI folders containing tiled images or embeddings, and (2) a folder named 'task-settings' containing task configurations and label files.

\subsection{Model Blocks}

\subsubsection{Backbone Builder for ROI} 
At the ROI level, we included various baseline backbones, organized into three categories: CNN-based backbones, Transformer-based backbones, and Hybrid backbones. The CNN-based category encompasses VGG16, VGG19 \cite{vgg}, ResNet50 \cite{resnet}, Xception \cite{xception}, and MobileNetV3 \cite{mobilevitv3}. The Transformer-based category also referred to as patch-learning, includes ViT-base \cite{vit} and Swin-base \cite{swin}. Meanwhile, Conformer \cite{conformer}, Crossformer \cite{crossformer}, and ResNet50-ViT \cite{msht} represent Hybrid backbones.

Beyond models pretrained on natural images, several foundational models trained on pathology datasets are also incorporated, such as ViT(UNI \cite{uni}), ViT(Virchow1 \cite{virchow}), ViT(Virchow2 \cite{virchow2}), ViT(Gigapath \cite{gigapath}), ViT(PuzzleTuning \cite{puzzletuning}), CONCH \cite{conch} and MUSK \cite{musk}. Unless otherwise specified, each model’s implementation is adapted from its official release with minimal structural modifications.

\subsubsection{Backbone Builder for WSI} 
Scaling up to whole slide images, we designed the three model-base with MIL-based methods, Transformer-based methods, and pre-trained foundational models. The MIL-based category encompasses baseline Average and Max pooling methods of SlideAve, SlideMax, and ABMIL \cite{abmil}, DSMIL \cite{dsmil}, CLAM \cite{clam}, and DTFD \cite{dtfd} for bag sampling design. The Transformer-based category includes TransMIL \cite{transmil}, SetMIL \cite{setmil}, and Graph-Transformer (GTP) \cite{gtp}. Lastly, the foundational slide-level models include pre-trained Gigapath \cite{gigapath}. Unless otherwise specified, each model is adapted from its official release with minimal implementation changes, primarily focusing on framework modifications to be called as modules.

\subsection{Task Blocks}

\subsubsection{Pre-training Modules}

Based on the pre-training task configuration, the backbone model and corresponding pre-training modules are selected. UnPuzzle currently supports self-supervised learning modules for pre-training, including BYOL \cite{byol}, DINO \cite{dinov2}, and MoCo-V3 \cite{mocov3} for contrastive learning, as well as MAE \cite{mae}, DINO V2, and SimMIM \cite{simmim} for masked image modeling. For ROI-level backbones, the training loop processes image inputs, while for WSI-level backbones, the loop processes embedded features, as illustrated in Figure~\ref{fig:fig_large}.f. Additionally, UnPuzzle accommodates multi-modal training pipelines such as CLIP \cite{clip}, PLIP \cite{plip} and CoCa \cite{coca}.

\subsubsection{Classification, Regression, and Multi-task Learning}
In UnPuzzle’s downstream phase, we offer a modular pipeline that dynamically constructs task-specific models using appropriate task heads and model backbones. Fully-connected layers for classification and regression are instantiated automatically based on task configurations. In multi-task learning scenarios, separate task heads are generated for each task. The training loop is configured accordingly for classification, regression, and multi-task learning, enabling seamless deployment in different downstream applications.

\subsubsection{Multi-modal Tasks}
For multi-modal downstream tasks, our pipeline covers both Vision Question Answering (VQA) training and Chain-of-Thought (CoT) inference. In particular, we have adapted LLava-med \cite{llavamed} and Coca \cite{coca} for these downstream tasks. An evaluation pipeline is also provided, allowing for comprehensive benchmarking against different backbone modules. As shown in Figure~\ref{fig:fig_large}.f, we integrated multi-modal capability for both slide-level and tile-level models.

\section{Experiments}

In this section, we outline the evaluation for the UnPuzzle framework, assessing its performance across a diverse set of several state-of-the-art foundation models and task-specific models for tile and slide-level downstream tasks. The full-scale distribution of the collected WSI and ROI datasets by organ is shown in Figure~\ref{fig:fig_large}.d. Their detailed benchmarking report, covering over 163 datasets and 200 tasks, will be available on our website, offering an in-depth analysis of benchmarking. In this paper, we demonstrate UnPuzzle framework with 9 WSI and 9 ROI datasets out of 163 collected public datasets for simplification.

\subsection{Demonstration Datasets and Tasks}

The demonstration datasets span a wide variety of organs and tasks, ensuring a comprehensive and robust evaluation of model performance. 
Regarding the demonstration datasets, they are formalin-fixed paraffin-embedded (FFPE) and frozen H\&E samples primarily sourced from repositories such as Genomic Data Commons (GDC), the Clinical Proteomic Tumor Analysis Consortium (CPTAC) and various public challenges. For WSI datasets, we have selected TCGA-BLCA for tumor staging, TCGA-BRCA for IHC-HER2 prediction, TCGA-CESC for grading, TCGA-Lung for non-small cell lung cancer subtyping, TCGA-MESO for histological diagnosis classification, TCGA-UCEC for overall survival (months) prediction, TCGA-UCS for tumor invasion percentage prediction, TCGA-UVM for tumor thickness prediction, and Camelyon16 for breast metastasis classification \cite{gdc_tcga,cam16}. 
For ROI datasets, we have selected NCT-CRC-HE-100k for colorectal cancer tissue classification \cite{nct-crc-he-100k}, PCam for breast metastasis detection \cite{pcam}, WBC for blood cells classification \cite{wbc}, GasHisSDB for gastric cancer classification \cite{gashissdb}, TCGA-MSI for colorectal cancer microsatellite instability screening, SIPakMed for cervical cancer cell types classification \cite{sipakmed}, OsteoTumor for osteosarcoma tumor classification \cite{osteotumor}, LC25000-Lung for non-small cell lung cancer subtyping, and LC25000-Colon for colon adenocarcinoma classification \cite{lc25000}. All the datasets are patient-wise split into train, validate and test subsets for benchmarking with a ratio of 7:1:2 or following the official split if available.

Regarding the tasks, we showcase several cancer detection, cancer subtyping, bio-marker prediction tasks: (1) tumor staging (Staging); (2) IHC-HER2 prediction (IHC-HER2); (3) grading; (4) non-small cell lung cancer subtyping (NSCLC); (5) histological diagnosis classification (HistDx); (6) overall survival (months) prediction (OSMonth); (7) tumor invasion percentage prediction (Inv); (8) tumor thickness prediction (Thick); (9) breast metastasis classification (BreastMet); (10) colorectal cancer tissue classification (CRCTissue); (11) blood cells classification (BlCells); (12) gastric cancer classification (GastricCa); (13) colorectal cancer microsatellite instability screening (CRC-MSI); (14) cervical cancer cell types classification (CerviCaTyping); (15) osteosarcoma tumor classification (OsteoTumor); (16) colon adenocarcinoma classification (ColonAdeno).  

\subsection{Demonstration Models} 
Our objective is to showcase UnPuzzle's scalability, flexibility, and benchmarking capabilities in supporting the evaluation of a wide range of pathology image analysis models. Several models can be applied as individual modules for various tasks, as shown in Figure~\ref{fig:fig_large}.e. For tile-level models, we have selected ViT-large (UNI weight), ViT-huge (Virchow2 weight), ViT-huge (ImageNet weight) \cite{vit}, and ResNet101 \cite{resnet} for evaluation on ROI datasets.
For slide-level models, after applying gigapath tile-embedding, we selected ABMIL \cite{abmil}, DSMIL \cite{dsmil}, CLAM \cite{clam}, TransMIL \cite{transmil}, LongNet \cite{gigapath} for evaluation on WSI datasets.

For benchmarking foundational models, we selected MUSK \cite{musk}, PuzzleTuning \cite{puzzletuning}, UNI \cite{uni}, CONCH \cite{conch}, Gigapath \cite{gigapath}, Virchow2 \cite{virchow2}, and the simple baseline method ResNet50 \cite{resnet} as the tile-level embedding model. After performing feature embedding, we evaluate the performance of downstream slide-level tasks using LongNet \cite{gigapath} for the Gigapath-embedded dataset and ABMIL \cite{abmil} for the remaining foundation models.

\subsection{Implementation Details} 
Our experiments were conducted on a GPU computing cluster, leveraging 4 nodes, each equipped with 4 NVIDIA RTX 4090 GPUs (24GB memory per GPU). The software environment included CUDA 12.4, PyTorch 2.4, and Python 3.10, ensuring compatibility with the latest frameworks and libraries. All experiments were executed under identical conditions to maintain consistency.

\subsection{Results}

\begin{table*}[ht]
    \centering
    \renewcommand{\arraystretch}{1.2}
    \tiny
    \caption{Left table: performance comparison of different slide-level models across various pathology tasks. In the upper section of the table, classification tasks are evaluated using accuracy, whereas the lower section assesses regression tasks using correlation coefficients. Right table: performance comparison of tile-level models across different datasets. UNI refers to a ViT-large model pretrained on the UNI dataset, Virchow2 denotes a ViT-huge model pretrained on the Virchow2 dataset, and ViT-h represents a ViT-huge model pretrained on ImageNet. The evaluation metric is accuracy.}
    \begin{tabular}{lccccc}
        \hline
        WSI Task & ABMIL & CLAM & DSMIL & LongNet & TransMIL \\
        \hline
        Staging & \textbf{60.20} & 52.30 & 58.00 & 51.10 & 45.50 \\
        HER2 & \textbf{62.80} & 54.10 & 59.00 & 52.50 & 59.00 \\
        Grade & \textbf{54.40} & \textbf{54.40} & 50.90 & \textbf{54.40} & 50.90 \\
        NSCLC & 75.70 & 68.60 & \textbf{76.30} & 69.80 & 73.40 \\
        HistDx & \textbf{60.00} & 46.70 & \textbf{60.00} & 53.30 & 53.30 \\
        BreastMet & 93.70 & 73.40 & 86.10 & \textbf{96.20} & 93.70 \\
        \hline
        OSMonth & 0.566 & 0.357 & \textbf{0.629} & 0.541 & 0.544 \\
        Inv & -0.205 & \textbf{0.456} & -0.232 & 0.227 & 0.390 \\
        Thick & 0.567 & -0.418 & 0.610 & \textbf{0.619} & 0.559 \\
        \hline
    \end{tabular}
    \quad
    \begin{tabular}{lcccc}
        \hline
        ROI Task & UNI & Virchow2 & ViT-h & ResNet101 \\
        \hline
        CRCTissue & \textbf{97.91} & 97.39 & 91.98 & 88.59 \\
        BreastMet & \textbf{92.56} & 65.46 & 75.44 & 72.41 \\
        BICells & 88.57 & \textbf{89.03} & 83.96 & 82.00 \\
        CRC-MSI & 81.37 & 81.01 & 61.88 & \textbf{82.72} \\
        GastricCA & 80.69 & 78.80 & \textbf{81.57} & 63.27 \\
        CerviCaTyping & 77.45 & 76.47 & \textbf{81.37} & 64.71 \\
        OsteoTumor & \textbf{85.71} & 84.82 & 83.93 & 79.46 \\
        NSCLC & \textbf{98.27} & 97.00 & 93.07 & 96.07 \\
        ColonAdeno & \textbf{96.60} & \textbf{96.60} & 96.00 & 94.10 \\
        \hline
    \end{tabular}
    \label{tab:model_performance_table}
\end{table*}

\begin{figure}[htbp]
  \centering
   \includegraphics[width=1.0\linewidth]{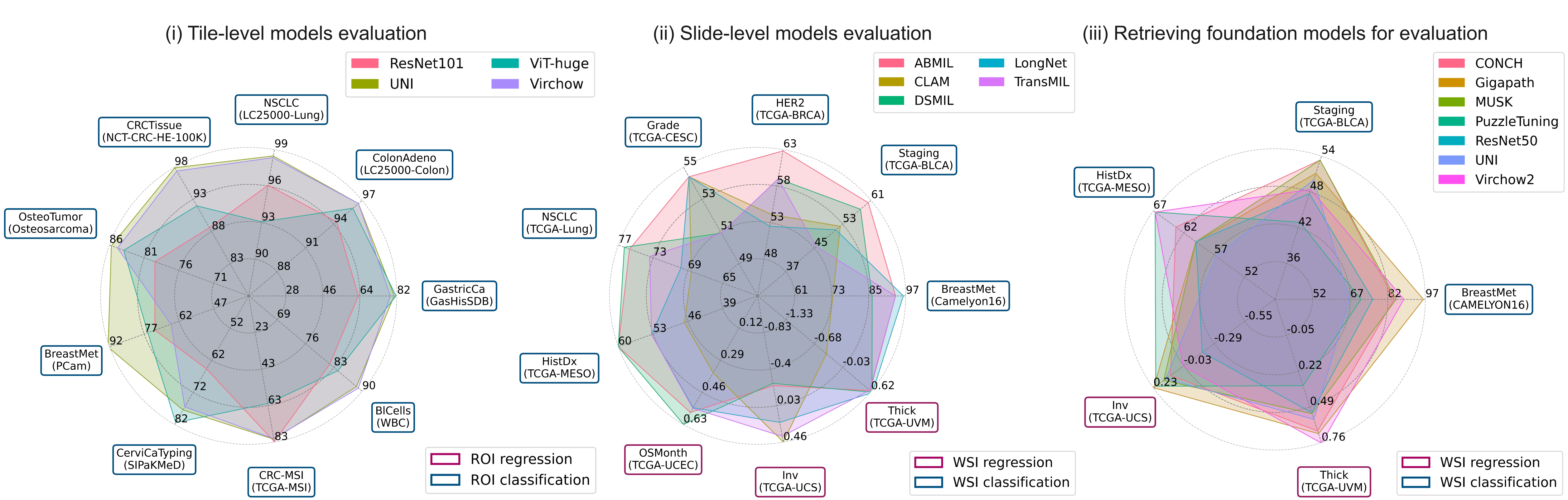}
   \caption{The evaluation results of slide-level and tile-level models among various downstream tasks.}
   \label{fig:fig_results}
\end{figure}

As shown briefly in Figure~\ref{fig:fig_results} and detailed in Table \ref{tab:model_performance_table}, UnPuzzle benchmarks comprehensive results. Due to space constraints, we summarize the key findings here and present the full details on our open-source website.

At the ROI level, across cancer detection and subtyping tasks—including NSCLC, CRCTissue, OsteoTumor, and ColonAdeno datasets—foundation models UNI and Virchow consistently outperform ImageNet-pretrained ResNet101 and ViT-huge in most downstream tasks. These results suggest that large-scale pretraining on domain-specific datasets builds better ROI models with a significant performance advantage.

However, when scaling up to WSI, the performance of complex models or training methods becomes less effective. In biomarker prediction tasks, ABMIL, a simpler model architecture, achieves superior performance in most cases. This highlights that straightforward designs can be effective for complex sequence modeling on WSI tiles, mitigating the risk of overfitting compared to more complex models. Furthermore, the performance of embedding models varies across tasks: Gigapath excels in Inv and BreastMet, Virchow2 leads in Thick, and MUSK achieves the best results in Staging. While larger-scale models generally produce better features, variations exist due to differences in feature distributions across downstream tasks.

Overall, the UnPuzzle benchmarks reveal that no single model consistently outperforms others. However, scaling up the pretraining of ROI models contributes to enhanced performance in both ROI and subsequent WSI modeling. These findings underscore the need for streamlined analysis workflows, where standardized performance comparisons play a critical role in optimizing model development.

\section{Conclusion}

In this paper, we introduce UnPuzzle, the first open-source, unified framework for pathological image analysis. With a modular design that standardizes workflows, UnPuzzle provides a flexible and scalable solution for pathology research. Through extensive benchmarking experiments across multiple datasets, we demonstrate the importance of standardized pipelines in fostering reproducibility, improving accessibility, and lowering the entry barrier for pathology AI research. By open-sourcing the pipeline and benchmark, we invite the research community to contribute to and adopt UnPuzzle, driving continuous improvements and innovations in the pathology field.

\bibliographystyle{splncs04}

\begin{thebibliography}{8}


\bibitem{song2023artificial}
Song, A.H., et al.: Artificial intelligence for digital and computational pathology. Nature Reviews Bioengineering \textbf{1}(11), 930--949 (2023)

\bibitem{mtl}
Zhang, Y., et al.: An overview of multi-task learning. National Science Review \textbf{5}(1), 30--43 (2018)


\bibitem{vgg}
Simonyan, K., et al.: Very Deep Convolutional Networks for Large-Scale Image Recognition. arXiv preprint arXiv:1409.1556 (2015), \url{https://arxiv.org/abs/1409.1556}

\bibitem{resnet}
He, K., et al.: Deep Residual Learning for Image Recognition. arXiv preprint arXiv:1512.03385 (2015)

\bibitem{vit}
Dosovitskiy., et al.: An Image is Worth 16x16 Words: Transformers for Image Recognition at Scale. International Conference on Learning Representations (2021), \url{https://openreview.net/forum?id=YicbFdNTTy}

\bibitem{xception}
Chollet, F.: Xception: Deep Learning with Depthwise Separable Convolutions. arXiv preprint arXiv:1610.02357 (2017), \url{https://arxiv.org/abs/1610.02357}

\bibitem{mobilevitv3}
Wadekar, S.N., et al.: MobileViTv3: Mobile-Friendly Vision Transformer with Simple and Effective Fusion of Local, Global and Input Features. arXiv preprint arXiv:2209.15159 (2022), \url{https://arxiv.org/abs/2209.15159}

\bibitem{swin}
Liu, Z., et al.: Swin Transformer: Hierarchical Vision Transformer using Shifted Windows. In: Proceedings of the 2021 IEEE/CVF International Conference on Computer Vision (ICCV), pp. 9992--10002 (2021), doi: 10.1109/ICCV48922.2021.00986

\bibitem{conformer}
Gulati, A., et al.: Conformer: Convolution-augmented Transformer for Speech Recognition. arXiv preprint arXiv:2005.08100 (2020), \url{https://arxiv.org/abs/2005.08100}

\bibitem{crossformer}
Wang, W., et al.: CrossFormer: A Versatile Vision Transformer Hinging on Cross-scale Attention. arXiv preprint arXiv:2108.00154 (2021), \url{https://arxiv.org/abs/2108.00154}

\bibitem{msht}
Zhang, T., et al.: MSHT: Multi-Stage Hybrid Transformer for the ROSE Image Analysis of Pancreatic Cancer. IEEE Journal of Biomedical and Health Informatics, \textbf{27}(4), 1946--1957 (2023), doi: 10.1109/JBHI.2023.3234289



\bibitem{clam}
Lu, M.Y., et al.: Data-efficient and weakly supervised computational pathology on whole-slide images. Nature Biomedical Engineering \textbf{5}, 555--570 (2021), doi: 10.1038/s41551-020-00682-w

\bibitem{dtfd}  
Zhang, H., et al.: DTFD-MIL: Double-tier feature distillation multiple instance learning for histopathology whole slide image classification. IEEE/CVF Conf. Comput. Vis. Pattern Recognit. (CVPR), 18780--18790 (2022)  

\bibitem{abmil}  
Ilse, M., et al.: Attention-based deep multiple instance learning. arXiv preprint arXiv:1802.04712 (2018) 

\bibitem{dsmil}  
Li, B., et al.: Dual-stream multiple instance learning network for whole slide image classification with self-supervised contrastive learning. IEEE/CVF Conf. Comput. Vis. Pattern Recognit. (CVPR), 14313--14323 (2021)  

\bibitem{transmil}  
Shao, Z., et al.: TransMIL: Transformer based correlated multiple instance learning for whole slide image classification. Adv. Neural Inf. Process. Syst. \textbf{34}, 2136--2147 (2021)  

\bibitem{setmil}  
Zhao, Y., et al.: SETMIL: Spatial encoding transformer-based multiple instance learning for pathological image analysis. In: Wang, L., Dou, Q., Fletcher, P.T., Speidel, S., Li, S. (eds.) MICCAI 2022. Lect. Notes Comput. Sci., vol. \textbf{13432}. Springer, Cham (2022)  

\bibitem{gtp}  
Zheng, Y., et al.: A graph-transformer for whole slide image classification. IEEE Trans. Med. Imaging \textbf{41}(11), 3003--3015 (2022)  

\bibitem{purity}  
Oner, M. U., et al.: Obtaining spatially resolved tumor purity maps using deep multiple instance learning in a pan-cancer study. Patterns (N Y) \textbf{3}(2), 100399 (2021), https://doi.org/10.1016/j.patter.2021.100399  



\bibitem{puzzletuning}
Zhang, T., et al.: PuzzleTuning: Explicitly Bridge Pathological and Natural Image with Puzzles. arXiv preprint arXiv:2311.06712 (2024), \url{https://arxiv.org/abs/2311.06712}

\bibitem{virchow}
Vorontsov, E., et al.: Virchow: A Million-Slide Digital Pathology Foundation Model. (Jan. 17, 2024), arXiv preprint arXiv:2309.07778, accessed: Feb. 29, 2024. [Online]. Available: \url{http://arxiv.org/abs/2309.07778}

\bibitem{virchow2}
Zimmermann, E., et al.: Virchow2: Scaling Self-Supervised Mixed Magnification Models in Pathology. (Nov. 06, 2024), arXiv preprint arXiv:2408.00738, doi: 10.48550/arXiv.2408.00738

\bibitem{uni}
Chen, R.J., et al.: Towards a general-purpose foundation model for computational pathology. Nature Medicine \textbf{30}(3), 850--862 (2024), doi: 10.1038/s41591-024-02857-3

\bibitem{gigapath}
Xu, H., et al.: A whole-slide foundation model for digital pathology from real-world data. Nature (May 2024), doi: 10.1038/s41586-024-07441-w

\bibitem{conch}  
Lu, M.Y., et al.: A visual-language foundation model for computational pathology. Nat. Med. \textbf{30}, 863--874 (2024)  

\bibitem{musk}
Xiang, J., et al.: A vision–language foundation model for precision oncology. Nature \textbf{638}, 769–778 (2025). https://doi.org/10.1038/s41586-024-08378-w


\bibitem{mae}
He, K., et al.: Masked Autoencoders Are Scalable Vision Learners. In: \textit{Proceedings of the 2022 IEEE/CVF Conference on Computer Vision and Pattern Recognition (CVPR)}, pp. 15979--15988 (2022), doi: 10.1109/CVPR52688.2022.01553

\bibitem{mocov3}  
Chen, X., et al.: An empirical study of training self-supervised vision transformers. IEEE/CVF Int. Conf. Comput. Vis. (ICCV), 9620--9629 (2021)  

\bibitem{dinov2}
Oquab, M., et al.: DINOv2: Learning Robust Visual Features without Supervision. \textit{arXiv preprint arXiv:2304.07193} (2024), \url{https://arxiv.org/abs/2304.07193}

\bibitem{byol}
Grill, J.B., et al.: Bootstrap your own latent: A new approach to self-supervised learning. In: \textit{Proceedings of the 34th International Conference on Neural Information Processing Systems (NeurIPS)}, pp. 1786–1799 (2020)

\bibitem{simmim}  
Xie, Z., et al.: SimMIM: a simple framework for masked image modeling. IEEE/CVF Conf. Comput. Vis. Pattern Recognit. (CVPR), 9643--9653 (2022)  

\bibitem{clip}  
Radford, A., et al.: Learning transferable visual models from natural language supervision. arXiv preprint arXiv:2103.00020 (2021)  

\bibitem{plip}  
Huang, Z., et al.: A visual–language foundation model for pathology image analysis using medical Twitter. Nat. Med. \textbf{29}, 1--10 (2023)  

\bibitem{coca}   
Yu, J., et al.: CoCa: contrastive captioners are image-text foundation models. arXiv preprint arXiv:2205.01917 (2022)  


\bibitem{llavamed}  
Li, C., et al.: LLaVA-med: training a large language-and-vision assistant for biomedicine in one day. Proc. 37th Int. Conf. Neural Inf. Process. Syst. (NIPS '23), 1240 (2023)  

\bibitem{cam16}
Bejnordi, B. E., et al.: Diagnostic assessment of deep learning algorithms for detection of lymph node metastases in women with breast cancer. JAMA \textbf{318}(22), 2199--2210 (2017)

\bibitem{gdc_tcga}  
Weinstein, J. N., et al.: The Cancer Genome Atlas Pan-Cancer analysis project. Nat. Genet. \textbf{45}, 1113--1120 (2013)  

\bibitem{nct-crc-he-100k}
Höhn, J., et al.: Colorectal cancer risk stratification on histological slides based on survival curves predicted by deep learning. NPJ Precis. Oncol. \textbf{7}, 98 (2023)  

\bibitem{pcam}  
Veeling, B. S., et al.: Rotation equivariant CNNs for digital pathology. In: Medical Image Computing and Computer Assisted Intervention–MICCAI 2018: 21st Int. Conf., Granada, Spain, Sept. 16–20, 2018, Proc. Part II, 210--218. Springer (2018)  

\bibitem{wbc}  
Kouzehkanan, Z. M., et al.: A large dataset of white blood cells containing cell locations and types, along with segmented nuclei and cytoplasm. Sci. Rep. \textbf{12}(1), (2022)  

\bibitem{gashissdb}  
Hu, W., et al.: GasHisSDB: A new gastric histopathology image dataset for computer aided diagnosis of gastric cancer. arXiv, cs.CV (2021), https://arxiv.org/abs/2106.02473  

\bibitem{sipakmed}  
Plissiti, M. E., et al.: Sipakmed: A new dataset for feature and image based classification of normal and pathological cervical cells in pap smear images. In: 2018 25th IEEE Int. Conf. Image Process. (ICIP), 3144--3148 (2018)  

\bibitem{osteotumor}  
Leavey, P., et al.: Osteosarcoma data from UT Southwestern/UT Dallas for viable and necrotic tumor assessment (Osteosarcoma-Tumor-Assessment) [Data set]. The Cancer Imaging Archive, https://doi.org/10.7937/tcia.2019.bvhjhdas (2019)  

\bibitem{lc25000}  
Borkowski, A. A., et al.: Mastorides, S. M.: Lung and colon cancer histopathological image dataset (LC25000). arXiv:1912.12142v1 [eess.IV] (2019)  





\end{thebibliography}

\end{document}